\documentclass[aps,twocolumn,superscriptaddress,preprintnumbers]{revtex4}
\usepackage{epsfig}
\usepackage{graphicx,amssymb,epsfig,amsmath}

\newcommand{\be}{\begin{eqnarray}}
\newcommand{\ee}{\end{eqnarray}}

\begin{document}

\title{Semi-device independent random number expansion protocol with $n\to 1$ quantum random access codes}
\author{Hong-Wei Li}
\affiliation{Key Laboratory of Quantum Information,University of
Science and Technology of China,Hefei, 230026,
 China}
\author{Marcin Paw{\l}owski}
\affiliation{ Department of Mathematics, University of Bristol,
Bristol BS8 1TW,United Kingdom}
\author{Zhen-Qiang Yin}
\affiliation{Key Laboratory of Quantum Information,University of
Science and Technology of China,Hefei, 230026,
 China}
\author{Guang-Can Guo}
\affiliation{Key Laboratory of Quantum Information,University of
Science and Technology of China,Hefei, 230026,
 China}
 \author{Zheng-Fu Han}
\affiliation{Key Laboratory of Quantum Information,University of
Science and Technology of China,Hefei, 230026,
 China}
\date{March 6, 2012}

\begin{abstract}
We study random number expansion protocols based on the $n \to 1$
quantum random access codes (QRACs). We consider them in the
semi-device independent scenario where the inner workings of the
devices are unknown to us but we can certify the dimensions of the
systems being communicated. This approach does not require the use
of the entanglement and makes the physical realization of these
protocols much easier than in the standard device independent
scenario. In our work, we propose a protocol for randomness
expansion, compute min-entropy for the semi-device independent
protocol, and investigate $n\to 1$ QRACs with a view to their use in
randomness expansion protocols. We also calculate the dependence of
the effectiveness of the randomness generation on $n$ and find it
optimal for $n=3$, and provide the explanation for this fact.

\end{abstract}
\maketitle

{\bf Introduction - } To certify that the given set of random
numbers is truly random is not an easy task. Since these numbers
cannot be generated by the deterministic algorithms, the devices
that generate them must operate according to some intrinsically
random physical process. The problem reduces then to the
certification that the device really performs the way it is supposed
to, at least within reasonable limits. The device independent
approach \cite{DI} allows the parties to establish the parameters
necessary for this certification without having to physically
examine the details of the device. This approach has been highly
successful in the quantum cryptography \cite{DIqkd1,DIqkd2,DIqkd3}
and recently has been taken to study also the randomness generation.
Colbeck \cite{Colbeck 1,Colbeck 2} has proposed the true random
number expansion protocol based on the GHZ test and Pironio {\it et
al}. \cite{Pironio 1} have proposed the protocol based on the Bell
inequality violations. All these protocols require entanglement
which has the negative effect on the complexity of the devices and
the rates of the randomness generation \cite{Pironio 1}.

The notion of semi-device independent, which assumes the knowledge
of the dimension of the underlying physical system but otherwise
nothing about the actual physical implementation of the
measurements, was first introduced by Liang {\it et al}.
\cite{liang} in the context of bounding entanglement. Working within
the same framework, a compromise between the need to know the
devices and the requirement for quantum resources was proposed
\cite{Witness 2}. In this scenario, secure key distribution without
entanglement has been proposed. More recently, the same approach has
also been used to certify true randomness \cite{Li}.

In both papers \cite{Witness 2,Li} the parameter estimated to
certify the devices was the average success probability of the $2
\to 1$ Quantum Random Access Code (QRAC). This protocol has been
firstly proposed in \cite{RAC1} allows one party, the sender, to
encode two classical bits in one qubit in such a way that the second
party which receives this qubit can decode any one of the two
classical bits with the success probability strictly greater than
$\frac{1}{2}$. Generalizations of it to $n \to 1$ QRACs, i.e. the
protocols where $n$ classical bits are encoded in a single qubit,
have then been proposed and studied \cite{RAC2,RAC3,RAC4,RAC5}. Here
we generalize the results from \cite{Li} and study the true random
number generation protocols based on $n \to 1$ QRACs. We are
particulary interested in the amount of randomness generated as a
function of $n$. We find that, remarkably, this function is not
monotonic and reaches the maximum for $n=3$. We provide the
explanation of this fact.

The paper is structured as follows. First we describe the semi-device independent scenario. Then we show how it can be used for the randomness generation. Later we calculate the amount of randomness generated as the function of $n$ and present in more detail the protocol that generates most randomness. We end with the discussion of our results.

{\bf Semi-device independent scenario - } The semi-device
independent random number generation \footnote{Since the protocol in
both device independent and semi-device independent cases require
some randomness for the certification procedure they are in fact
randomness expansion rather than generation protocols.} protocol
requires two black boxes, which are used for the state preparation
and measurement respectively. \\\\
1. State preparation black box: The preparation box is given
randomly one of the $2^n$ inputs $a$ represented by $n$ different
bits $a=a_1\cdot\cdot\cdot a_n$. For each input this box emits the
state $\rho_{a}$ which is sent to the measuring box.\\ \\2. State
measurement black box: There one of $n$ different measurements
$y=\{1,\cdot\cdot\cdot,n\}$, is chosen and the outcome $b=\{0,1\}$
returned.\\\\
3. QRAC average success probability estimation: Applying random
input numbers $a,y$ and measurement outcomes $b$ to estimate QRAC
average success probability, which can be used for guarantee
randomness of the measurement outcomes.
\\\\4. Random number generation: The parameter that is used to the
randomness of the measurement outcome $b$ conditioned on the input
values $a$ and $y$ is the min-entropy function (It's used for
extractors both in classical and quantum cases)
\cite{renner1,renner2} given by
\begin{equation}
\begin{array}{lll}
H_{\infty}(B|A,Y)\equiv-log_{2}[max_{b,a,y}P(b|a,y)].
\end{array}
\end{equation}

 Since it is the semi-device independent scenario we do not
have any knowledge of the measurements and the preparations apart
from the fact that for all $a$ the states $\rho_a$ are of dimension
2, and they are not parts of any lager entangled systems. This
scenario is schematically depicted on FIG. 1.

\begin{figure}[!h]\center
\resizebox{8cm}{!}{
\includegraphics{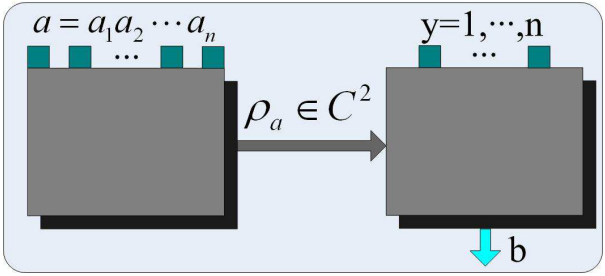}}
\caption{Semi-device independent random number expansion protocol.}
\end{figure}

{\bf Certification parameters -}
The certification of the device requires the estimation of the probabilities
\begin{equation}
\begin{array}{lll}
P(b|ay)=tr(\rho_{a}M_y^b),
\end{array}
\end{equation}
where, $M_y^b$ is the measurement operator acting on the two
dimensional Hilbert space with the input parameter $y$ and the
output parameter $b$. The advantage of the semi-device independent
approach is that these probabilities are the only numbers that need
to be calculated in order to find the parameter that certifies the
device. This parameter is, as we have mentioned, the average success
probability of the $n \to 1$ QRAC given by \be
S_n\equiv\frac{1}{n2^{n}}\sum_{a,y} P(b=a_y|a,y), \ee but for our
purposes it will be more convenient to use
\begin{equation}
\begin{array}{lll}                 \label{TTT}
T_n\equiv \Sigma_{a,y}(-1)^{a_y}E_{ay},
\end{array}
\end{equation}
where $E_{ay}=P(b=0|ay)$.

By $T_n^{classical}$ we denote the greatest value of $T_n$ possible if the communicated system is a classical bit. By $T_n^{quantum}$ the same value for qubit. The corresponding values for $n=2,3,4,5$ are
\begin{equation}
\begin{array}{lll}                 \label{TTTT}
T_2^{classical} \leq~2,~~T_2^{quantum} \leq~2.828427,~~n=2,\\
T_3^{classical} \leq~6,~~T_3^{quantum} \leq~6.928203,~~n=3,\\
T_4^{classical} \leq12,~~T_4^{quantum} \leq15.454813,~~n=4,\\
T_5^{classical} \leq30,~~T_5^{quantum} \leq34.172467,~~n=5.\\
\end{array}
\end{equation}
They correspond to the codes presented in \cite{RAC4}.

The intuition behind using these codes to generate the randomness is that the more bits are encoded into the single qubit the less certain is
 the correct guessing of any of them. We find the result that the amount of randomness is not a monotonous function of $n$ remarkable since the parameters $S_n$ and $\frac{T_n^{quantum}}{T_n^{classical}}$ are.

{\bf Amount of randomness - }
The main result of this paper is the amount of the randomness generated by $n \to 1$ QRACs in the semi-device independent scenario.

More precisely, we solve the following optimization problem:
\begin{equation}
\begin{array}{lll}
\text{minimize:}\quad  max_{b,a,y}P(b|a,y)\\\\
\text{subject to:}\quad E_{ay}=tr(\rho_{a}M_y^0)\\
~~~~~~~~~~~~~~~~~~~\Sigma_{a,y}(-1)^{a_y}E_{ay}=T_n
\end{array}
\end{equation}
where the optimization is carried over arbitrary quantum states
$\rho_{a}$ and measurement operators
$\{M_1^0,\cdot\cdot\cdot,M_n^0\}$ defined over 2-dimensional Hilbert
space. In the most general case, we should consider the positive
operator valued measure (POVM) $\{M_j^0,M_j^1\}$, where
 $M_j^0+M_j^1=I$ for $j\in\{1,\cdot\cdot\cdot,n\}$.
Fortunately, Masanes \cite{Masanes} has proved that only the
projective measurements should be considered in the case of
2-measurement outcomes have to be considered( Jordan {\it et al}.
\cite{Jordan,Hanggi} have showed that such binary measurements
commute with a projective measurement). Since $T$ is the linear
expression of the probabilities, we can only consider pure states
\cite{Witness 1} in the numerical calculation. Without the loss of
generality, the state preparation and measurements in our numerical
analysis can be illustrated with the following equations

\begin{equation}
\begin{array}{lll}
\rho_{a}=|\varphi(a)\rangle\langle\varphi(a)|,
\end{array}
\end{equation}

\begin{equation}
\begin{array}{lll}
|\varphi(a)\rangle= \left(\begin{array}{ccc}
cos (\frac{\theta_{a}}{2})\\
e^{i\eta_{a}}sin (\frac{\theta_{a}}{2})
\end{array}\right),
\end{array}
\end{equation}

\begin{equation}
\begin{array}{lll}
M_1^0= \left(\begin{array}{ccc}
1&0\\
0&0
\end{array}\right),
\end{array}
\end{equation}

\begin{equation}
\begin{array}{lll}
M_j^0= \left(\begin{array}{ccc}
cos^2(\frac{\psi_j}{2})&\frac{1}{2}e^{-i\omega_j} sin(\psi_j) \\
\frac{1}{2}e^{i\omega_j} sin(\psi_j)&sin^2(\frac{\psi_j}{2})
\end{array}\right),
\end{array}
\end{equation}
where
$a\in\{\overset{2^n}{\overbrace{\underset{n}{\underbrace{00\cdot\cdot\cdot0}},\cdot\cdot\cdot,\underset{n}{\underbrace{11\cdot\cdot\cdot1}}}}\}$,
$j\in\{\overset{n-1}{\overbrace{2,\cdot\cdot\cdot,n}\}}$,
$0\leq\theta_{a},\psi_j\leq\pi$, $0\leq\eta_{a},\omega_j\leq2\pi$.

By solving the minimization problem, we get the min-entropy bound of
the measurement outcome for the given $n\to 1$ QRAC. If we set
$T_n=T_n^{quantum}$ we get the maximal amount of randomness
generated by the given QRAC. We obtain the following results
\be\nonumber n=2 \quad H_\infty\simeq0.2284 \\
\nonumber n=3 \quad H_\infty\simeq0.3425 \\
\nonumber n=4 \quad H_\infty\simeq0.1388 \\
\nonumber n=5 \quad H_\infty\simeq0.1024\ee

 Note that the min-entropy
bound based on $2 \to 1 $ QRAC is larger than our previous result
\cite{Li}, the reason for which is that the maximal quantum
correlation value $T_2^{quantum}$ is approximated to six decimal
places. Fig. 2 shows that even the small changes of $T_2$ lead to very large
changes in the min-entropy bound if $T_2$ is close to the maximal quantum value.

\begin{figure}[!h]\center
\resizebox{8cm}{!}{
\includegraphics{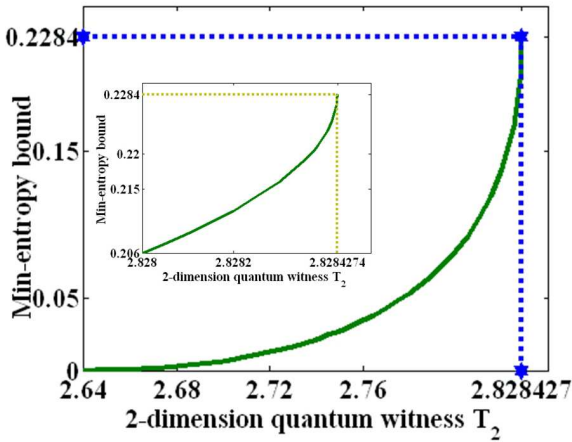}}
\caption{The relationship between the average efficiency of $2 \to
1$ QRAC measured by $T_2$ and the min-entropy bound.}
\end{figure}

As we have already mentioned the largest amount of the randomness is
generated with the $3\to 1$ QRAC. Now we describe in this code more
detail.

{\bf Randomness generated by $3 \to 1 $ QRAC - } The optimal
protocol generates 0.1141 more per the run of the experiment than
the first semi-device independent randomness generation protocol
introduced in \cite{Li}. The lower bound on the amount of randomness
generated as the function of $T_3$ is plotted in Fig. 3. The result
show that we can get the positive amount of randomness as soon as
$T_3>6.65$.

\begin{figure}[!h]\center
\resizebox{8cm}{!}{
\includegraphics{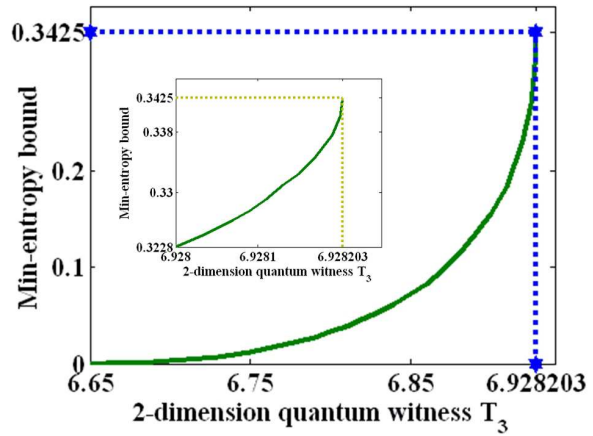}}
\caption{The relationship between the average efficiency of $3 \to
1$ QRAC measured by $T_3$ and the min-entropy bound. We see that the
 positive amount of randomness can be generated as soon as $T_3>6.65$, and the
maximal value of the min-entropy is 0.3425 with $T_3=6.928203$. }
\end{figure}

The parameters $T_3^{quantum}$ and $H_\infty=0.3425$ can be obtained
using the following $3\to 1$ QRAC. The state prepared by the first
black box is given by
\begin{equation}
\begin{array}{lll}
|\varphi(000)\rangle=cos(\xi)|0\rangle+e^{i\frac{\pi}{4}}sin(\xi)|1\rangle,\\
|\varphi(001)\rangle=cos(\xi)|0\rangle+e^{-i\frac{\pi}{4}}sin(\xi)|1\rangle,\\
|\varphi(010)\rangle=cos(\xi)|0\rangle+e^{i\frac{3\pi}{4}}sin(\xi)|1\rangle,\\
|\varphi(011)\rangle=cos(\xi)|0\rangle+e^{-i\frac{3\pi}{4}}sin(\xi)|1\rangle,\\
|\varphi(100)\rangle=sin(\xi)|0\rangle+e^{i\frac{\pi}{4}}cos(\xi)|1\rangle,\\
|\varphi(101)\rangle=sin(\xi)|0\rangle+e^{-i\frac{\pi}{4}}cos(\xi)|1\rangle,\\
|\varphi(110)\rangle=sin(\xi)|0\rangle+e^{i\frac{3\pi}{4}}cos(\xi)|1\rangle,\\
|\varphi(111)\rangle=sin(\xi)|0\rangle+e^{-i\frac{3\pi}{4}}cos(\xi)|1\rangle,
\end{array}
\end{equation}\\
where $\xi=arccos\sqrt{\frac{1}{2}+\frac{\sqrt{3}}{6}}$. For the
state measurement in the second black box, we use the three
projective measurements with the following bases
\begin{equation}
\begin{array}{lll}
\{M_1^0=|0\rangle\langle0|,~~~~M_1^1=|1\rangle\langle1|\},\\
\{M_2^0=|0'\rangle\langle0'|,~~M_2^1=|1'\rangle\langle1'|\},\\
\{M_3^0=|0''\rangle\langle0''|,~~M_3^1=|1''\rangle\langle1''|\},
\end{array}
\end{equation}
where $|0'\rangle=\frac{1}{\sqrt{2}}(|0\rangle+|1\rangle),
|1'\rangle=-\frac{1}{\sqrt{2}}(|0\rangle-|1\rangle)$,
$|0''\rangle=\frac{1}{\sqrt{2}}(|0\rangle+i|1\rangle),
|1''\rangle=-\frac{1}{\sqrt{2}}(|0\rangle-i|1\rangle)$.

{\bf Discussion - } We have presented the family of randomness generation protocols based on the $n \to 1$ QRACs. The certification that the device indeed produces truly random numbers is done in the semi-device independent scenario, which combines the advantages of the standard device independent approach with less requirements on the resources required. We have found that the optimal member of the family, in the sense of the amount of the randomness generated, is the $3\to 1$ QRAC.

The remarkable feature of this family of codes is that the amount of the randomness as a function of $n$ is not monotonic. To understand why it is so, consider the Bloch sphere. Both the states and the measurements can be represented by the unit vectors on it. If one looks at the constructions for the QRACs from \cite{RAC4} one notices that the states are almost evenly spread on the whole surface of the sphere. Therefore, as $n$ grows, for any measurement the nearest state gets closer and closer. This however does not happen for the transition from $n=2$ to $n=3$. The reason for this is that the states and the measurements for the optimal QRAC lie in one plane and do not use the full size of the space.

The $3 \to 1$ QRAC is the most efficient semi-device independent
randomness generation protocol known. It remains the open question
if even better protocols exist.

In comparison to fully device-independent random number expansion
protocols, the protocols presented here are much easier to
realize, which makes them a good object for future studies, and one of the
most important aspects of it should be the full security proof.

{\bf Acknowledgements - }
 H-W. L. thanks Y-C. W., X-B. Z. and Y. Y. for helpful discussions. H-W.L., Z-Q.Y., G-C.G. and Z-F.H. are supported by the
National Basic Research Program of China (Grants No. 2011CBA00200
 and No. 2011CB921200), National Natural Science Foundation of China (Grant NO. 60921091), and National High Technology Research and Development Program of China (863 program) (Grant No.
 2009AA01A349)
. M.P. is supported by UK EPSRC.
 To whom correspondence should be addressed, Email:
$^a$maymp@bristol.ac.uk, $^b$zfhan@ustc.edu.cn.

\end{document}